\documentclass{elsart}
\usepackage{times}
\usepackage{graphicx}
\usepackage{natbib}
\usepackage{amssymb}

\def\oiii{[O\,{\sc iii}]}
\def\halpha{\hbox{H$\alpha$}}
\def\hbeta{\hbox{H$\beta$}}
\def\ni{\hbox{[N\,{\sc i}]}}
\def\nii{\hbox{[N\,{\sc ii}]}}

\newcommand{\sauron}{{\texttt {SAURON}}}

\newcommand{\oasis}{{\texttt {OASIS}}}

\begin{document}
\runauthor{Jes\'us Falc\'on-Barroso}
\begin{frontmatter}

\title{Morphology and kinematics of the ionised gas in early-type 
galaxies}

\author[Sterrewacht]{Jes\'us Falc\'on-Barroso\thanksref{Email}},
\author[Oxford]{Marc Sarzi},
\author[CRAL]{Roland Bacon},
\author[Oxford]{Martin Bureau},
\author[Sterrewacht]{Michele Cappellari},
\author[Oxford]{Roger L. Davies},
\author[CRAL]{Eric Emsellem},
\author[Roch]{Kambiz Fathi},
\author[Oxford]{Davor Krajnovi\'c},
\author[ESA]{Harald Kuntschner},
\author[Sterrewacht]{Richard M. McDermid},
\author[Groningen]{Reynier F. Peletier},
\author[Sterrewacht]{Tim de Zeeuw}

\address[Sterrewacht]{Sterrewacht Leiden, Niels Bohrweg~2, 2333~CA
Leiden, The Netherlands}
\address[Oxford]{Denys Wilkinson Building, University of Oxford, Keble 
Road, Oxford, United Kingdom}
\address[CRAL]{CRAL, 9~Avenue Charles Andr\'e, 69561, Saint Genis
Laval,
France}
\address[Roch]{RIT, 85 Lomb Memorial Drive, Rochester, New York 14623,
USA}
\address[ESA]{Space Telescope European Coordinating Facility, European 
Southern Observatory, Karl-Schwarzschild-Str~2, 85748 Garching, Germany}
\address[Groningen]{Kapteyn Astronomical Institute, P.O. Box 800, 9700 
AV Groningen, The Netherlands}
\thanks[Email]{jfalcon@strw.leidenuniv.nl}

\begin{abstract}
We present results of our ongoing study of the morphology and kinematics
of the ionised gas in 48 representative nearby elliptical and lenticular
galaxies using the \sauron\ integral-field spectrograph on the 4.2m
William Herschel Telescope. Making use of a recently developed
technique, emission is detected in 75\% of the galaxies. The
ionised-gas distributions display varied morphologies, ranging from
regular gas disks to filamentary structures. Additionally, the
emission-line kinematic maps show, in general, regular motions with
smooth variations in kinematic position angle. In most of the galaxies,
the ionised-gas kinematics is decoupled from the stellar counterpart,
but only some of them present signatures of recent accretion of gaseous
material. The presence of dust is very common in our sample and is
usually accompanied by gas emission. Our analysis of the \oiii/\hbeta\
emission-line ratios, both across the whole sample as well as within
the individual galaxies, suggests that there is no unique mechanism
triggering the ionisation of the gas.
\end{abstract}
\end{frontmatter}
\typeout{SET RUN AUTHOR to \@runauthor}

\section{Introduction}
The advent of integral-field spectrographs has opened a whole new window
of possibilities for extra-galactic studies. Coupled with large
telescopes and advances in adaptive optics systems (AO), integral-field
units (IFUs) offer a unique opportunity to go further in the exploration
of astrophysical phenomena earlier in the history of the universe and at
smaller scales. It is therefore not surprising that IFUs are quickly
becoming standard instruments on all the 8m-class telescopes. Besides
the efforts to investigate the formation and evolution of galaxies in
the early universe, much can be learned by studying in detail the {\em
fossil\/} record of nearby galaxies using IFUs on intermediate-size
telescopes. The \sauron\ project (de Zeeuw et al. 2002), is a first step 
in this direction. The project is designed to perform a systematic study 
of a representative sample of early-type, nearby galaxies using the \sauron\
(Bacon et al. 2001) and \oasis\ (Bacon et al. 1995; McDermid et al. 2004) 
IFUs at the 4.2m William Herschel Telescope. Early results of the survey have 
revealed a variety of structures much richer than usually recognised in 
early-type galaxies (de Zeeuw et al. 2002; McDermid et al. 2004; Emsellem 
et al. 2004; see also R. McDermid's contribution in this workshop).

Here we focus on the distribution and kinematics of the ionised gas of a
subset of galaxies to illustrate the diversity of structures and
ionisation processes present in the \sauron\ representative sample of 48
elliptical and lenticular galaxies. The sample contains an equal
fraction of elliptical and lenticular galaxies, with equal number of
field and cluster galaxies in each morphological group. We will make use
of the two-dimensional information delivered by \sauron\ to compare the
distribution and kinematics of the stars and gas and to study the nature
and origin of the ionised gas in these systems. A more extended analysis
of the properties of the ionised gas for the entire sample is presented
in Sarzi et al. (2005). Furthermore, the study of the stellar and gas
properties of the 24 Sa bulges in the \sauron\ sample is addressed in
Falc\'on-Barroso et al. (2005).

\section{Extraction of the ionised gas}
\label{sec:method}
Early-type galaxies were traditionally thought to be uniform stellar
systems with little or no gas and dust. However, a considerable number
of imaging and spectroscopy studies have changed this view (Sadler \&
Gerhard 1985; van Dokkum \& Franx 1995; Goudfrooij 1999; Tran et al 2001).
In the spectral range delivered by \sauron\ there are three potential 
emission lines that can be measured (i.e. \hbeta, \oiii$\lambda\lambda$4959, 
5007, \ni$\lambda\lambda$5198,5200). Given that the stellar contribution to
the overall spectrum in our sample of galaxies remains significant, it
is important to perform an accurate continuum subtraction in order to
determine reliably the fluxes and kinematics of the emission lines. We
investigated several methods to perform this task and settled on a
procedure that {\em simultaneously} fits both the stellar and gaseous
contributions to derive the fluxes, velocities and velocity dispersions
of the emission lines.\looseness-1 

Traditional methods to separate the stars and gas consisted of the
subtraction of the best matching stellar template, in the regions free
of emission, to the full spectrum of the galaxy. As described in
Sarzi et al. (2005), this methodology can significantly overestimate the
flux and velocity dispersion of the measured emission lines. Extensive
simulations show that the new procedure is much more accurate and
appears to be superior to previous ones, allowing us to detect emission
down to an equivalent width of 0.1 \AA.

\section{Distribution of the ionised gas}
\label{sec:distrib}
The distribution of the ionised gas in our sample shows diverse
morphologies. In general the distribution follows that of the stars,
although there are several cases where the situation is much more
complex (i.e. filaments, rings, spiral arms, lanes). 

The incidence of ionised-gas emission in our sample is 75\%. Seven
galaxies show only weak traces of emission, and there are five cases
with no detection. By morphological type, lenticular galaxies display a
slightly higher content of ionised gas than elliptical galaxies (83\%
versus 66\%). Similar percentages were found when the sample was divided
according to environment (83\% field, 66\% cluster). One important
remark is that the fraction of galaxies with clearly detected emission
in the Virgo cluster drops to only 55\% (10/18), with just 3/9
ellipticals showing the presence of gas. The incidence of ionised-gas
shows no correlation with either luminosity or the presence of a bar in
the galaxy. The detection rates found for our sample are in good
agreement with those of Macchetto et al. (1996), who found emission in 
85\% of the lenticular and 68\% of the ellipticals in their sample.

In Figure~\ref{fig2}, we present maps of the stellar and ionised-gas
distribution and kinematics for four of the 48 galaxies in our sample.
NGC\,2974 reveals a fast rotating gas disk co-rotating with respect to
the stars. Despite the regular appearance of the ionised-gas
distribution, the equivalent width (EW) map of the \oiii\ line (see
Fig.~\ref{fig2}) highlights the presence of an inner ring and two spiral
arms that extend all the way out to the limits of our field-of-view (see
Krajnovi\' c et al. 2005a for a detailed analysis of this galaxy using 
\sauron\ data). A similarly regular case is that of NGC\,4526, although 
in this case the ionised gas is confined to a well defined, fast rotating 
disk in the central kpc of the galaxy. NGC\,2768 is a well-known galaxy where
the distribution of the ionised-gas appears to be perpendicular to that
of its stellar counterpart (i.e. polar-ring, Fried \& Illingworth 1994;
Bertola et al. 1992). The \sauron\ \oiii\ EW map shows that the gas 
distribution has a filamentary morphology along the galaxy minor axis. 
The gas distribution in NGC\,4278 displays a peculiar integral-sign 
pattern that is closely followed by the gas velocity field. The stellar 
and gas kinematics appear to be misaligned by increasingly wider angles, 
as they twist in opposite directions towards the outer parts of the 
field-of-view.

\section{What powers the observed nebular emission?}
\label{sec:ratios}
Diagnostic diagrams of \oiii/\hbeta\ vs \nii/\halpha\ (Veilleux \& 
Osterbrock 1987; Kauffmann et al. 2003) have been used extensively in 
the past to investigate the trigger of the ionisation of the gas (i.e. AGN, 
star-formation). Many scenarios have been invoked to explain the presence 
of ionised gas in early-type galaxies: central AGN, hot ($10^7$K) gas 
(Sparks et al. 1989; de Jong et al. 1990), young stars (Shields 1992), 
post-AGB stars (Binette et al. 1994), or shocks (Dopita \& Sutherland 1995). 
Within the wavelength range delivered by \sauron\ only the
\hbeta\ and \oiii\ lines can be used for this purpose. Despite this
limitation, the \oiii/\hbeta\ ratio serves as a good indicator to
locate regions where emission is due to young stars and also to trace
variations of the ionisation mechanisms within a single galaxy.

Regions displaying low \oiii/\hbeta\ ratios (i.e. $\le$1) are usually
interpreted as indicative of star-formation (Ho et al. 1997). However,
higher ratios (i.e. $>$1) could still indicate star-formation if the
metallicity of the gas is sufficiently high (Veilleux \& Osterbrock 1987). 
Additionally other mechanisms can lead to high ratios. Since it is unlikely 
that the metallicity of the ionised gas varies abruptly within galaxies 
(unless there has been some recent accretion of material), large variations
across maps of the \oiii/\hbeta\ ratio are more likely to be produced
by changes in the ionisation mechanism rather than changes in the metallicity.

Figure~\ref{fig2} shows the \oiii/\hbeta\ ratios for the galaxies
discussed in Section~\ref{sec:distrib}. The \oiii/\hbeta\ ratio maps
clearly reveal the presence of radial gradients and substructures
in the ionisation properties of the gas in the four galaxies. The ratio
is moderately high in NGC\,2768, NGC\,2974, and NGC\,4278, while it is
very low in NGC\,4526. In this last galaxy, the location of the low
\oiii/\hbeta\ values corresponds to the location of a prominent dust
disk (as shown in the unsharp-masked image), and suggests ongoing star
formation. In this respect, there seems to be a clear link between the
ionised structures found in the \oiii/\hbeta\ maps with those seen in
the \oiii\ equivalent width and kinematic maps (i.e. ring in NGC\,2974,
or integral-sign pattern in NGC\,4278).

The remaining sample of 44 galaxies also displays a great diversity of
\oiii/\hbeta\ ratios within galaxies, but also across the sample. This
suggests that either there are many mechanisms at play in the ionisation
of the gas, that the metallicity of the ionised gas is very
heterogeneous in those galaxies, or both.

\section{Relation between gas and dust}
\label{sec:dust}
In order to study the dust distribution and compare it with that of
the ionised gas we have generated unsharp-masked images using archival
{\it Hubble Space Telescope} images. We found that, consistent with
previous results, dust generally follows the ionised-gas distribution
(Goudfrooij et al. 1994; Tomita et el. 2000; Tran et al. 2001). The 
opposite situation is not always true, as we have found galaxies with 
clear presence of emission with no traces of dust in the unsharp-masked 
images. As found by previous authors (Ho et al. 2002), regular dust 
distributions generally are associated with smooth velocity fields, 
however the lack of regular dust lanes does not imply the presence of 
irregular kinematics.

The four cases presented in Figure~\ref{fig2} display a wide range of
dust morphologies that correlate in different degrees with the
structures seen in the distribution of the ionised gas. The prominent
dust disk in NGC\,4526 traces closely the distribution of the gas. The
ring-like structure and spiral arms in NGC\,2974 are also followed by
dusty structures at the same locations. The dust morphology is less
defined in NGC\,2768 and NGC\,4278, but still suggests an orientation
for the dust that is similar to that of the gas distribution.

\begin{figure*}
\begin{center}
  \includegraphics[width=1.0\linewidth]{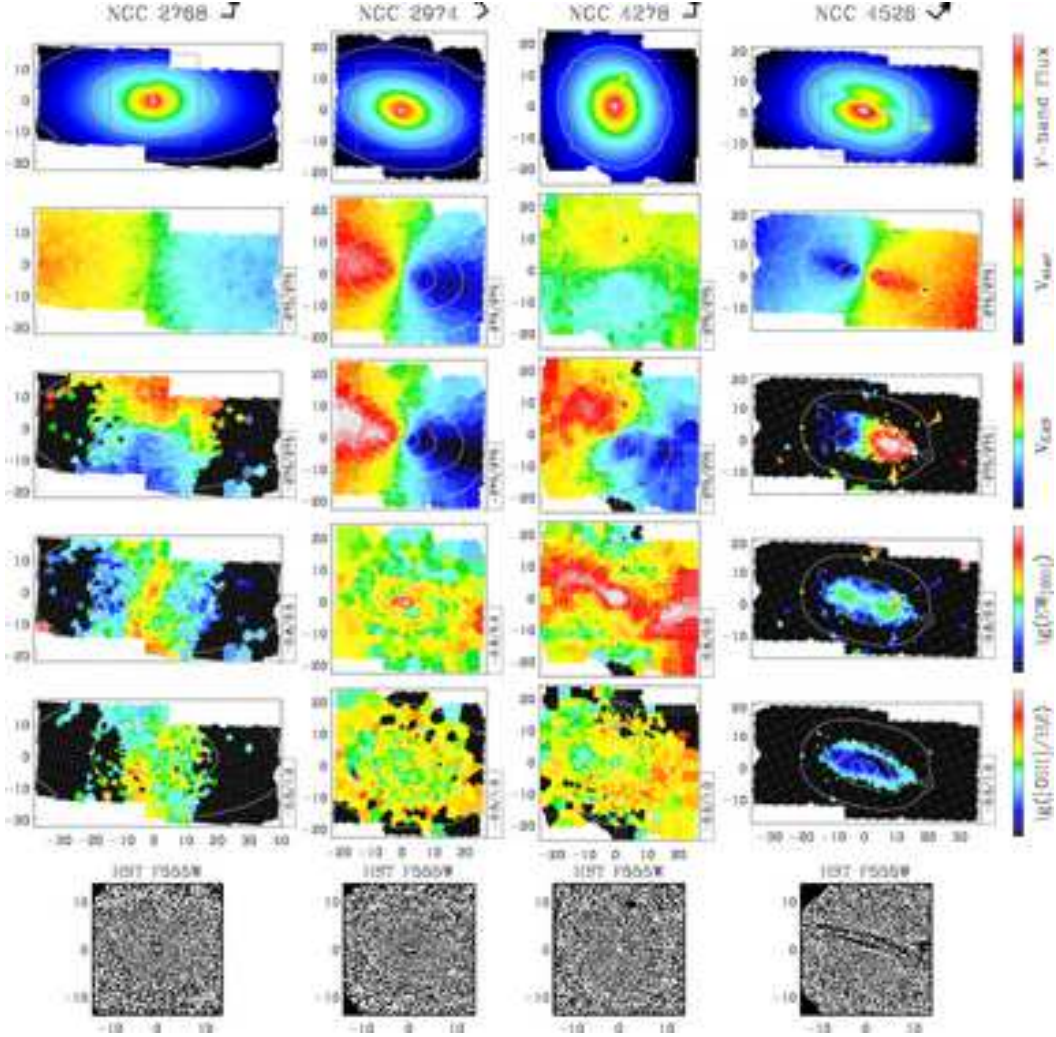}
\end{center}
\caption{\sauron\ maps for 4 galaxies representative of the gas
properties in the 48 E and S0 galaxies surveyed by the \sauron\ project
(Sarzi et al. 2005). From top to bottom: i) the reconstructed total
intensity, ii) the stellar velocity, iii) the ionised-gas velocity,  iv)
the equivalent width of the \oiii\ emission line in \AA\ (in log units),
v) the value of the \oiii/\hbeta\ ratio (also in log units), and vi)
unsharp-masked images obtained from HST observations. The cuts levels
are indicated in the box at the right-hand side of each map. The grey
boxes on the top figures indicate the field-of-view of the HST images in
the last row. Notice how the gas distribution can vary from galaxy to
galaxy, and how the kinematic major axis of the ionised gas deviates
from that of the stars. The \oiii/\hbeta\ ratio maps clearly reveal the
presence of radial gradients and substructures in the ionization
properties of the gas. The low \oiii/\hbeta\ values corresponding to the
dust disk in NGC\,4526 suggest ongoing star formation.}
\label{fig1}
\end{figure*}

\section{The Origin of the ionised gas}
\label{sec:origin}
The measurement of misalignments between the kinematics of the
gaseous and stellar components in early-type galaxies has often been
used to determine the relative importance of accretion events and the
internal production of gas through stellar mass-loss (e.g. Bertola et 
al. 1992). The orientation of the dust relative to that of the stars has 
also served this purpose (e.g. van Dokkum \& Franx 1995). In order
to quantify the presence of decoupled gaseous components in our sample
and to investigate their dependence on environment, we have measured the
mean misalignment between the stellar and ionised gas using {\it
kinemetry}, a generalisation of surface photometry to the higher-order
moments of the line-of-sight velocity distribution of galaxies
(Krajnovi\' c et al. 2005b). 

\begin{figure*}
\begin{center}
  \includegraphics[width=0.95\linewidth]{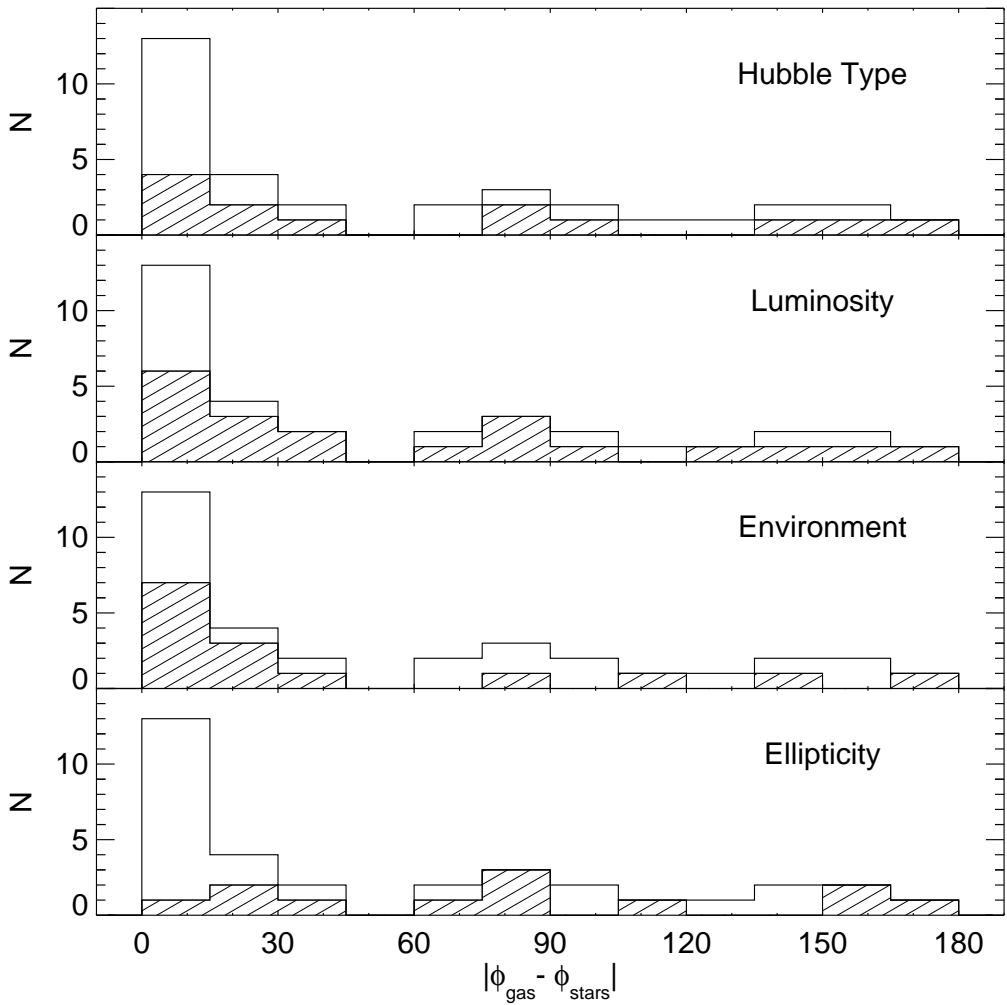}
\end{center}
\caption{Distribution of misalignments between the kinematic stellar
and gaseous axes. In all panels the solid line represents the
distribution of misalignments for the 48 E and S0 galaxies. The shaded
areas in the different panels represent (from top to bottom):
ellipticals, galaxies brighter than M$^{*}_{B}=-19.7$ mag \citep{eep88},
field galaxies, and roundest galaxies ($\epsilon \leq 0.2$).}
\label{fig2}
\end{figure*}

In Figure~\ref{fig2} (solid line) we show the distribution of the
average misalignments for the galaxies in our sample of 48 galaxies with
sufficiently extended emission. The distribution of values appears to be
skewed towards small misalignments. However, the fact that there is
still a significant fraction of galaxies showing mild to strong
misalignments suggests that the ionised gas has not a purely internal
origin (in which only small misalignments are expected). The overall
shape of the distribution remains unchanged when the sample is divided
into elliptical and lenticular galaxies. No dependency is found on
neither environment nor galaxy luminosity. We do, however, find a strong
dependence on apparent flattening of the galaxy (see Fig.~\ref{fig2},
bottom panel). It appears that the distribution of misalignments of the
most flattened objects accounts for the observed excess of co-rotating
over counter-rotating gas and stellar systems in the whole sample. The
roundest objects ($\epsilon<$0.2) display a more uniform distribution.
This result suggests that rotational support might be important to
explain the observed dependence.

\section{Concluding remarks}
We have presented some examples of the morphological and kinematical
state of the ionised gas in a representative sample of early-type
galaxies using data from the \sauron\ spectrograph at the 4.2m William
Herschel Telescope. The results of our analysis reveal a wide range of
morphological, but also kinematical structures. It also shows that
there must be several mechanisms responsible for the ionisation of the
gas, not only across galaxies, but also within galaxies. The presence of
dust is common in our sample and is usually accompanied by gas emission.
Finally, we studied the distribution of misalignments between the stars
and gas in our sample, and conclude that the origin of the ionised gas
cannot be purely internal or external. \looseness-1

\ack
JFB acknowledges support from the Euro3D Research Training Network,
funded by the EC under contract HPRN-CT-2002-00305. This work is based
on observations obtained at the WHT on the island of La Palma, operated
by the Isaac Newton Group at the Observatorio del Roque de los Muchachos
of the Instituto de Astrof\'\i sica de Canarias. 


\end{document}